\documentclass[aps,prl,twocolumn,showpacs,superscriptaddress]{revtex4}
\usepackage{graphicx}

\begin{document}

\title{Emergence of Fermi pockets in an excitonic CDW melted novel superconductor}

\author{D. Qian}
\author{D. Hsieh}
\affiliation{Department of Physics, Joseph Henry Laboratories of
Physics, Princeton University, Princeton, NJ 08544}
\author{L. Wray}
\affiliation{Department of Physics, Joseph Henry Laboratories of
Physics, Princeton University, Princeton, NJ 08544}
\author{E. Morosan}
\affiliation{Department of Chemistry, Princeton University,
Princeton, NJ 08544}
\author{N.L. Wang}
\affiliation{Institute of Physics, Chinese Academy of Sciences,
Beijing 100080, China}
\author{Y. Xia}
\affiliation{Department of Physics, Joseph Henry Laboratories of
Physics, Princeton University, Princeton, NJ 08544}
\author{R.J. Cava}
\affiliation{Department of Chemistry, Princeton University,
Princeton, NJ 08544}

\author{M.Z. Hasan}
\affiliation{Department of Physics, Joseph Henry Laboratories of
Physics, Princeton University, Princeton, NJ 08544}
\begin{abstract}

A superconducting (SC) state (T$_c$ $\sim$ 4.2K) has very recently
been observed upon successful doping of the CDW ordered triangular
lattice TiSe$_2$, with copper. Using high resolution photoemission
spectroscopy we identify, for the first time, the momentum space
locations of the doped electrons that form the Fermi sea of the
parent superconductor. With doping, we find that the kinematic
nesting volume increases whereas the coherence of the CDW order
sharply drops. In the superconducting doping, we observe the
emergence of a large density of states in the form of a narrow
electron pocket near the \textit{L}-point of the Brillouin Zone with
\textit{d}-like character. The \textit{k}-space electron
distributions highlight the unconventional interplay of CDW to SC
cross-over achieved through non-magnetic copper doping.


\end{abstract}



\date{22 November, 2006}

\maketitle


Charge-density-waves (CDW) and superconductivity (SC) are two of the
most fundamental collective quantum phenomena in solids. The
competition between the two is one of the central issues of
condensed matter physics \cite{morosan, kivelson, palee, takada,
naxtas, moncton}. A microscopic study of this competition is often
difficult for two reasons : one, most systems involve some forms of
intervening phases such as the magnetic order observed in the parent
state of cuprate\cite{palee} or cobaltate\cite{takada}
superconductors; two, there lacks systematically tunable electronic
parameters compatible with microscopic \textit{k}-space imaging
techniques such as angle-resolved photoemission spectroscopy
(ARPES). Until very recently, no system existed where controlled
chemical tuning allowed this competition to be studied in
microscopic detail. Recent investigations have shown that upon
controlled intercalation of TiSe$_2$ with copper (Cu$_x$TiSe$_2$)
one can systematically tune between a CDW state and a novel
superconducting state which emerges near x = 0.04, with a maximum
T$_c$ of 4.15K reached near x = 0.08 \cite{morosan}. A quantum
melting of CDW order was also observed, recently, in the parent
compound by pressure \cite{snow}. Cu$_x$TiSe$_2$ thus provides a
rare opportunity to systematically investigate the \textit{quantum
phase transition from a CDW ordered state to superconductivity} in
microscopic detail.

The parent compound TiSe$_2$ has been known since the 1960s and has
been extensively studied \cite{wilson, disalvo, oldarpes, pillo,
rossnagel, kidd, cui} due to the unconventional nature of the CDW
state. In this Letter, we use high resolution ARPES to image the
momentum space evolution of doped electrons as they evolve from the
undoped CDW ordered phase to an optimal superconductor by copper
doping for the first time. A detailed knowledge of momentum space
distribution of electronic states is critically important in
developing a theory for the superconducting order parameter
(\textit{s}, \textit{p} or \textit{d}-wave) realized in this novel
materials class. Based on the emergent Fermi surface topology and
momentum-space distinctions of the participant electrons in each
phase we directly show that the superconducting phase emerges upon
the gradual loss of coherence of the CDW state on the same site;
thus the two states compete in a microscopic manner and not due to
phase separation or other extrinsic effects. In the superconducting
doping we observe the emergence of large density of states in the
form of an extremely narrow electron pocket of about 100 meV
bandwidth.

\begin{figure}[t]
\includegraphics[width = 8cm]{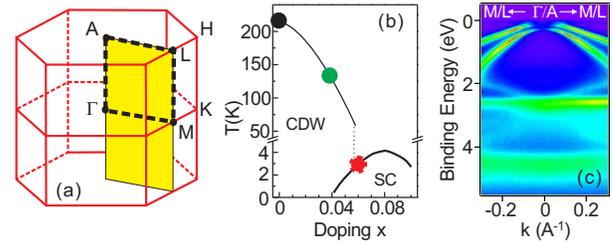}
\caption{{\bf{Sample characterization}} (a) Brillouin zone layout of
Cu$_x$TiSe$_2$ (b) Doping-temperature phase diagram suggests CDW is
a competing phase for superconductivity (c) High resolution valence
band profile of Cu$_x$TiSe$_2$ along the M/L-$\Gamma$/A-M/L momentum
space cut measured at 20K. The top two bands are the
$\Gamma^3$(A$^{3-}$) and $\Gamma^{2-}$ bands (in agreement with LDA
calculations \cite{lda}) characteristic of these compounds and
typically observed in samples with high degree of crystallinity. }
\end{figure}

\begin{figure*}[t]
\includegraphics[width = 12cm]{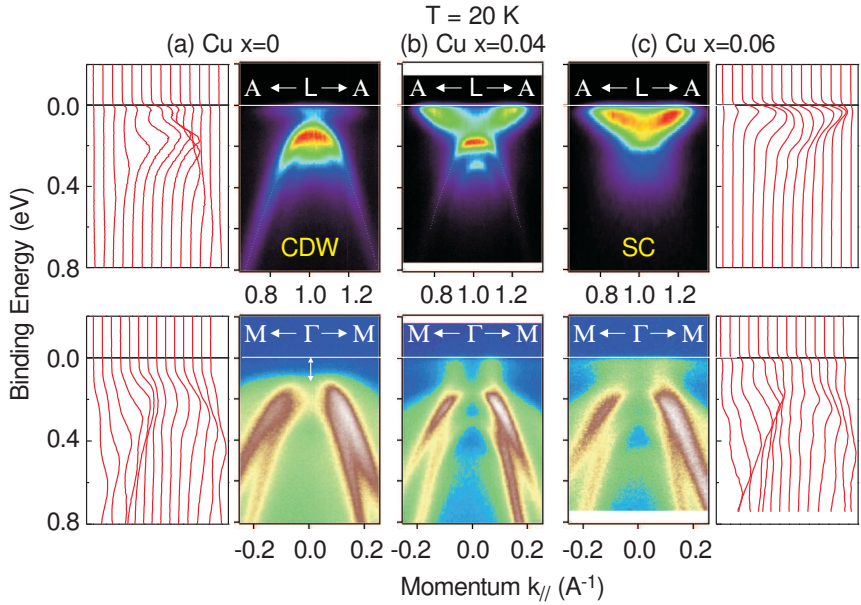}
\caption{{\bf{CDW state melts near superconducting doping :}} Low
temperature (T $<<$ T$_{CDW}$ at x=0) doping evolution of bands in
the vicinity of the \textit{L}-point (top row) and the $\Gamma$
points (bottom row). With increasing copper concentration the
intensity of the folded band decreases and spectral weight
concentrates near the \textit{L}-points in the form of an electron
pocket with bandwidth about 0.1 eV. (Bottom row) With increasing
doping spectral weight near the M-$\Gamma$-M cuts continues to shift
to lower energies and the deeper-lying bands tend to split further
from each other.}
\end{figure*}

\begin{figure}[t]
\includegraphics[width = 9cm]{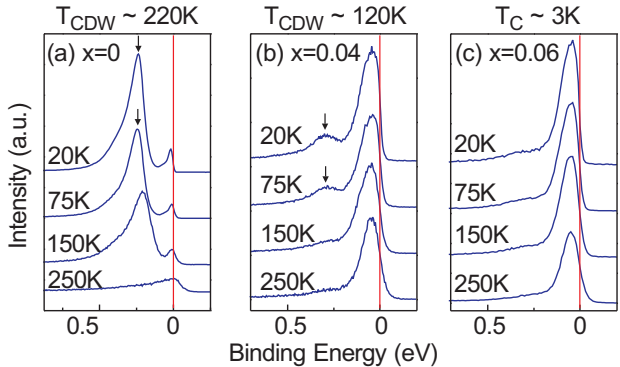}
\caption{{\bf{Temperature evolution of spectral function with copper
doping}} Energy distribution curves for k = L+0.1 \AA$^{-1}$ are
shown for x=0, 0.04 and 0.06 dopings. The peak at 0.25 eV in x=0 at
T=20K is due to the band folding. With increasing copper doping the
intensity of the 0.25 eV peak drops while the intensity of the
quasiparticle peak grows dramatically.}
\end{figure}

Undoped TiSe$_2$ is a layered material with trigonal structure and
hexagonal Brillouin zone. From previous high/finite temperature
photoemission studies of the undoped material \cite{oldarpes},
TiSe$_2$ is believed to be either a semimetal or a very small gap
semiconductor. It undergoes a (2x2x2) structural transition below
the CDW transition temperature T$_c$=220K. Upon copper doping, the
resistivity anomaly near 200K is suppressed and samples become more
metallic-like and eventually superconducts at low temperatures as
measured in powder samples \cite{morosan}. Recently, high quality
single crystal samples were grown \cite{hc2} making
\textit{k}-resolved measurements possible. ARPES measurements were
carried out at beamline ALS-12.0.1 at the Advanced Light Source and
SRC-U-NIM beamline at the Synchrotron Radiation Center. The data
were collected with 13eV to 79eV photons with better than 10meV to
25meV energy resolution and an angular resolution better than 1\% of
the Brillouin zone using Scienta analyzers with chamber pressure
better than 8x10$^{-11}$ torr. The $\Gamma$ and \textit{L} points
are determined by tracing the band maximum and minimum via changing
photon energy \cite{oldarpes}. Photon energies of 13eV and 56eV were
used for the measurements near the $\Gamma$-point whereas 19 and
78eV photons were used for the data collection near the \textit{L}
point. Cleaving the samples in situ at 20K or 250K resulted in shiny
flat surfaces characterized by diffraction to be clean and well
ordered with the same symmetry as the bulk. The measured valence
band (Figure-1) was found to be in good agreement with LDA band
theory \cite{lda}.

\begin{figure*}[t]
\includegraphics[width = 12cm]{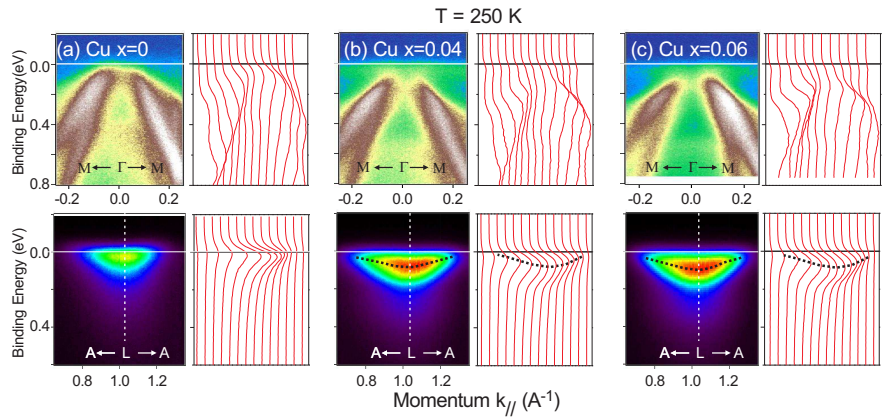}
\caption{{\bf{Electron pockets emerge and grow with doping :}} High
temperature (T $>>$ T$_{CDW}$) band evolution in the vicinity of the
\textit{L}-point. With increasing Cu concentration Fermi level rises
up. At the $\Gamma$ point this is seen as lowering of the Se 4p
bands (a-c, top row) and the emergence of low-energy states. At the
\textit{L}-point, doping gives rise to an electron pocket with
bandwidth of about 0.1 eV. The \textit{L}-pocket grows with
increasing copper doping.}
\end{figure*}

The low temperature evolution of momentum-resolved electronic
structure of TiSe$_2$ with copper doping is shown in Figure-2. A
spectacular change of low-energy electronic structure is observed
with systematic doping. The emergence of a pocket like band near the
\textit{L}-point is very prominent at low temperatures. With copper
doping the density of states near the \textit{L}-point increases
systematically while the states below 0.2 eV are seen to be
systematically suppressed. At low temperatures, the states below 0.2
eV near the \textit{L}-point are due to the folding of bands from
the $\Gamma$-point via the 2x2x2 reconstruction which is a
consequence of the CDW order. With copper doping, we observe a
weakening of the folding strengths in x=0.04 and x=0.06 samples. In
fact in the x=0.06 samples folding is extremely weak suggesting a
short range nature or near-absence of the CDW order. Figure-2(bottom
row) shows the evolution near the $\Gamma$-point. A systematic
sinking of the band multiplet is observed to occur with doping. This
is accompanied by some finite growth of intensity near the Fermi
level over an extended \textit{k}-range. However, no clear
quasiparticle crossing is resolved in the energy dispersion curves
near the $\Gamma$-point. Selected energy dispersion curves for the
temperature evolution of electronic states near \textit{k} =
\textit{L}+0.1 \AA$^{-1}$ with doping are shown in Figure-3. The
prominent peak at 0.25 eV in x=0 at T=20K is due to band folding and
constitutes the key spectral signature of CDW order. This peak grows
in intensity and moves to higher binding energies with decreasing
temperature. This behavior is consistent with an earlier report on
the x=0 samples\cite{rossnagel}. With increasing copper doping the
intensity of the 0.25 eV peak drops while the intensity or the
density of states near the Fermi level grows dramatically.

Above the CDW transition temperature (for T $>$ 220K), in the doped
system, we find that the chemical potential is near the mid-point of
the leading edge of the quasiparticle peak which is a clear
indication that the doped compounds are metals (Fig-2$\&$4). In
contrast, the location of the chemical potential in the undoped
system is at the peak (highest intensity) suggesting it is either
the thermal tail signal from a band that barely grazes the Fermi
level. Therefore, it is not directly resolved within the finite
temperature ARPES data. However, for data taken at sufficiently low
temperatures (T $\sim$ 20K) the x=0 parent compound is found to be
metallic (see Fig-2 and 3). The high temperature evolution of
electronic structure with copper doping is shown in Figure-4. With
doping, spectral weight is transferred toward the Fermi level,
however, no clear crossing is resolved near the $\Gamma$-point in
our study. In contrast, a pocket like band near the \textit{L}-point
is found to sink below the Fermi level with doping (bottom row). In
the band structure calculations, this band derives from the
Ti-\textit{3d} like states. A fit to the EDCs at k=\textit{L} for
x=0 sample suggests that the band barely touches the Fermi level or
slightly above the Fermi level as measured in high temperature. With
copper doping we observe a systematic sinking of this band reaching
about 0.1 eV in bandwidth near x=0.06 doping. Since the size of the
pocket grows with increasing doping we conclude an electron doping
(as opposed to hole doping) scenario with the intercalation of
copper. This is consistent with the negative sign of thermopower
(S/T) reported in the doped systems \cite{morosan}.

We have carried out a detailed measurement of electron distribution
\textit{n(k)} in the \textit{A-L} plane in all metallic samples of
Cu$_{x}$TiSe$_2$. The presented data in Figure-5 is shown with an
energy integration window of about 20 meV within the Fermi level and
was taken in the first quadrant then symmetrized over the Brillouin
zone. This is checked by direct measurements in other quadrants and
found to be consistent with the symmetrization procedure apart from
intensity variations due to changing matrix elements. Large
intensities in the form of ellipses are observed at the midpoints of
the hexagonal arms. In the superconducting doping samples,
Cu$_{0.06}$TiSe$_2$, these zone-boundary ellipses resolve as
electron-like pockets (Fig-5) which originate from the sinking band
near the \textit{L}-point in Fig-4(bottom row).

With increasing doping the gap closes and low-energy states cross
the Fermi level. This leads to the emergence of kinematic nesting
instabilities (non-2x2x2 type) along \textit{A-L} and \textit{A-H}.
Our data show that the CDW state loses long-range coherence with
doping but no new superstructure is observed as can be seen by band
folding. Such loss of coherence is quite evident from decreasing
intensity of the 0.25 eV feature near the \textit{L}-point (Fig-3).
Within band theory no nesting is expected to generate a 2x2x2
commensurate superstructure for the electron distribution in the x=0
compound \cite{lda}. With copper doping we find that the electron
pocket near \textit{L}-point continues to grow thereby increasing
the phase space for nesting but no new band folding is observed and
the strengths of 2x2x2 folding continues to drop ruling out nesting
as a mechanism for CDW in the doped system.

Following the theoretical suggestion of W. Kohn \cite{kohn}, an
electronic (excitonic-like) mechanism recently been proposed to
account for the CDW order of the parent compound \cite{kidd}(and
references there-in). If this scenario is correct, one would expect
the electron-hole coupling necessary to generate a gap to decrease
in the doped systems (which were previously unavailable for study)
since the CDW loses long-range coherence with doping. In this
scenario, the evidence for the loss of coherence comes from the fact
that the bottom of the pocket band at \textit{L}-point (Fig-2(c)/top
panel) is not as flat as seen in the undoped compound as reported in
ref-\cite{kidd}. It is unclear if this scenario is supported in the
doped systems since no clear recovery of the parabolic-like shape is
evident in the doped systems.

Irrespective of the CDW decoherence mechanism, it is well known that
increasing density of states (DOS) at the Fermi level,
\textit{D(E$_f$)}, enhances superconducting pairing correlations in
a system. The structurally analogous novel superconductor
Na$_x$CoO$_2$ exhibits enhancement of \textit{D(E$_f$)} with Na
doping in the presence of hydration \cite{takada}. It is known that
\textit{D(E$_f$)} also increases in the Fe-doped systems,
Fe$_x$TiSe$_2$ and Fe$_x$TiTe$_2$ \cite{disalvo, cui}, however no
superconductivity is observed in any of these systems. It is likely
that the magnetic moments in the Fe$_x$Ti(Se/Te)$_2$ are strong
enough to suppress superconductivity via the non-unitary
interactions (renormalizing T$_c$ to zero). In Cu$_x$TiSe$_2$,
copper is monovalent (Cu$^{+1}$) and non-magnetic (3d$^{10}$ and
S=0). Therefore its main role is to donate an electron to the Ti-Se
layer which then delocalizes. We have observed in this work that
this electron mostly enters the parent band around the
\textit{L}-point that has \textit{d}-like (Ti-\textit{3d}) character
which exhibits weak dispersion (less than 100 meV). Therefore, the
overall effect of non-magnetic copper doping is to raise the
\textit{d}-like density of states, \textit{D(E$_f$)}, at the Fermi
level. Such increases in \textit{D(E$_f$)} leads to a rise in T$_c$
in many systems, as expected in the BCS-Eliashberg scenario:
k$_B$T$_c$ = $\hbar$$\omega$$\exp^{-\frac{1}{D(Ef)V}}$.

\begin{figure}[t]
\includegraphics[width = 5.5cm]{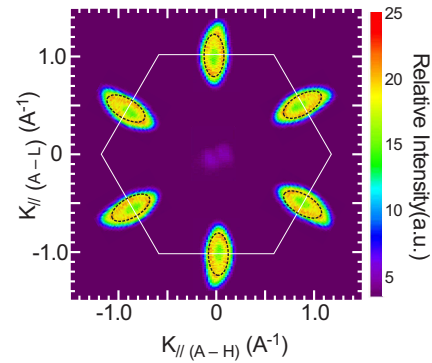}
\caption{{\bf{Normal state electron distribution of superconductor
Cu$_{0.06}$TiSe$_2$}} With increasing copper doping electrons near
the L-point form a narrow pocket and generate a six-fold symmetric
Fermi contour. A finite but weak intensity is also observed at the
$\Gamma$/A point. The presented summary of data above is measured in
the first quadrant at T=20K then symmetrized over the Brillouin
zone. The black dotted lines highlight the shape of the pockets.}
\end{figure}

Finally, we note that the Fermi velocity (v$_F$) of the
\textit{L}-point electron pocket is about 0.4 $\pm$ 0.1 eV.$\AA$
(see E vs. k in Fig-4(c)). This allows us to estimate the expected
the coherence length, $\xi$ $\sim$ 0.19 $\frac{\hbar v_F}{k_B T_c}$
$\sim$ 29 nm, under the assumption that this band is the Fermi sea
for the superconducting pairing in this compound. This value is
remarkably close to the value of 22 nm estimated from H$_{c2}$
measurements \cite{hc2} for x=0.07 samples which provides further
support for our identification of this \textit{L}-pocket electrons
taking part in the superconducting pairing in this compound. It is
known that the CDW phase of x=0 parent compound is associated with
the large softening of an \textit{L}-phonon \cite{holt}. It is
possible that the \textit{L}-phonon plays a key role in the
superconducting mechanism. Our preliminary results do not
necessarily rule out an electronic mechanism of superconductivity.
This would be a topic for our future studies.

In conclusion, we have reported the first measurement of the
low-lying states and the Fermi surface topology of the copper doped
CDW superconductor. Our results show that an electron-like pocket
near the \textit{L}-point continues to grow upon doping into the
superconducting phase. A systematic loss of the long-range coherence
of the CDW state is observed at low temperatures via the loss of
spectral intensity from the folded bands with increasing doping
suggesting that the CDW microscopically competes with
superconductivity. Momentum space distribution of electronic states
is crucial in developing a microscopic theory/mechanism for the
superconducting order parameter realized in this novel materials
class. The reported evolution of overall electronic structure with
copper doping will be further helpful in formulating a theory of
CDW-supercondctivity competition in general.

\begin{acknowledgments}

This work is primarily supported by DOE grant DE-FG-02-05ER46200.
RJC and MZH acknowledges partial support through NSF(DMR-0213706).
ALS is supported by the Office of Basic Energy Sciences
DOE/DE-AC02-05CH11231; SRC/UW is supported by NSF/DMR-0537588.

\end{acknowledgments}


\begin{thebibliography}[

\bibitem{morosan} E. Morosan \textit{et.al}, Nature Phys., \textbf{2}, 544 (2006).
\bibitem{kivelson} S.A. Kivelson \textit{et.al}, Nature, \textbf{393}, 550 (1998).
\bibitem{palee} P.A. Lee \emph{et.al.}, Rev. Mod. Phys. \textbf{78}, 17
(2006); S. Sachdev and S.C. Zhang Science \textbf{295}, 452 (2002);
M. Imada {\it et al.,} Rev. Mod. Phys. \textbf{70}, 1039 (1998).
\bibitem{takada} K. Takada \textit{et al.}, Nature \textbf{422}, 53
(2003); D. Qian \textit{et al.}, Phys. Rev. Lett. \textbf{96},
216405 (2006); D. Qian \textit{et al.}, Phys. Rev. Lett.
\textbf{96}, 046407 (2006), also Phys. Rev. Lett. \textbf{97},
186405 (2006).
\bibitem{naxtas} L. Fang \textit{et.al}, Phys. Rev. B \textbf{72}, 14534 (2005).
\bibitem{moncton} D.E. Moncton \textit{et.al}, Phys. Rev. Lett. \textbf{34},
734 (1975).
\bibitem{snow} C.S. Snow \textit{et.al}, Phys. Rev. Lett. \textbf{91}, 136402
(2003).
\bibitem{wilson} J.A. Wilson and A.D. Yoffe, Adv. Phys. \textbf{18}, 193 (1969).
\bibitem{disalvo} F.J. Di Salvo \textit{et.al}, Phys. Rev. B \textbf{14}, 4321 (1976).
\bibitem{oldarpes} R.Z. Bachrach \textit{et.al}, Phys. Rev. Lett. \textbf{37},
40 (1976); M.M. Traum \textit{et.al}, Phys. Rev. B \textbf{17},
1836(1978); G. Margaritondo \textit{et.al}, Phys. Rev. B
\textbf{23}, 3765 (1981); O. Anderson \textit{et.al}, Phys. Rev.
Lett. \textbf{55}, 2188 (1985); N.G. Stoffel \textit{et.al}, Phys.
Rev. B \textbf{31}, 8049 (1985).
\bibitem{pillo}Th. Pillo \textit{et.al}, Phys. Rev. B \textbf{61}, 16213 (2000).
\bibitem{rossnagel} K. Rossnagel \textit{et.al}, Phys. Rev. B \textbf{65}, 235101 (2002).
\bibitem{kidd} T.E. Kidd \textit{et.al}, Phys. Rev. Lett. \textbf{88}, 226402 (2002).
\bibitem{cui} K. Yamazaki \textit{et.al}, Physica B, \textbf{351}, 262 (2004); X.Y. Cui \textit{et.al}, Phys. Rev. B \textbf{73}, 085111 (2006).
\bibitem{lda} A. Zunger \textit{et.al}, Phys. Rev. B \textbf{17}, 1839 (1978).
\bibitem{hc2} E. Morosan \textit{et.al}, cond-mat/0611310 (2006).
\bibitem{kohn} W. Kohn, Phys. Rev. Lett. \textbf{19}, 439 (1967).
\bibitem{holt} M. Holt \textit{et.al}, Phys. Rev. Lett. \textbf{86}, 3799
(2001).

\end{thebibliography}
\end{document}